\documentclass[useAMS,usenatbib]{mn2e}

\usepackage{graphicx}
\usepackage{txfonts}
\usepackage{natbib}
\newcommand\aj{AJ}
\newcommand\apj{ApJ}
\newcommand\apjs{ApJS}
\newcommand\apss{Ap\&SS}
\newcommand\aap{A\&A}
\newcommand\mnras{MNRAS}
\newcommand\apjl{ApJ}
\newcommand\pasp{PASP}

\newcommand\araa{ARA\&A}

\title[]{Helium and Multiple Populations in the Massive Globular Cluster NGC\,6266 (M\,62)}
\author[A.\,P.\, Milone]
{A.\,P.\,Milone$^{1}$\\
$^{1}$Research School of Astronomy \& Astrophysics, Australian National University, Mt Stromlo Observatory, via Cotter Rd, Weston, ACT 2611
, Australia \\}

\begin{document}
\date{Draft Version Sep, 24, 2014}

\pagerange{\pageref{firstpage}--\pageref{lastpage}} \pubyear{2013}

\maketitle
\label{firstpage}

\begin{abstract}  
Recent studies suggest that the helium content of multiple stellar populations in globular clusters (GCs) is not uniform. The range of helium 
varies from cluster to cluster with  
more massive GCs having, preferentially, large helium spread. 
GCs with large helium variations also show extended-blue horizontal branch (HB).\\
I exploit {\it Hubble Space Telescope} photometry to investigate multiple stellar populations in NGC\,6266 and infer their relative helium abundance. This cluster is an ideal target to investigate the possible connection between helium, cluster mass, and HB morphology, as it exhibits an extended HB and is among the ten more luminous GCs in the Milky Way.\\
The analysis of color-magnitude diagrams from multi-wavelength photometry reveals 
that 
also
NGC\,6266, similarly to other massive GCs, hosts a 
double main sequence (MS), with the red and the blue component made up of the 79$\pm$1\% and the 21$\pm$1\% of 
stars, respectively. The red MS is consistent with a stellar population with primordial helium while the blue MS is highly helium-enhanced by $\Delta$Y=0.08$\pm$0.01.
 Furthermore, the red MS exhibits an intrinsic broadening that can not be attributed to photometric errors only and is consistent with a spread in helium of $\sim$0.025 dex.\\  
The comparison between NGC\,6266 and other GCs hosting helium-enriched stellar populations supports the presence of a correlation among helium variations, cluster mass, and HB extension.
\end{abstract}


\section{Introduction}\label{sec:intro}
Recent studies have shown that the CMD of a small but increasing number of GCs is made of two or more continuous sequences of stars which run from the bottom of the MS up to the red-giant branch (RGB) and correspond to stellar populations with different content of light elements and helium (e.g.\,Milone et al.\,2012a;   Piotto et al.\,2014 and references therein).
 The most-widely accepted possibility is that GCs have experienced at least two episodes of star formation, with second-generation star formed from material polluted by a previous stellar generation (e.g.\,Ventura et al.\,2001; Decressin et al.\,2007; De Mink et al.\,2009; Denissenkov \& Hartwick\,2014, but see Bastian et al.\,2013 and Cassisi \& Salaris\,2014 for an alternative scenario).

The knowledge of the helium content of multiple stellar populations  provides constraint on the nature of the polluters and is fundamental to shed light on the series of events that led from massive clouds in the early Universe to the GCs we see today. A correct determination of the stellar age, mass, and mass function requires accurate measure of Y.
Since the 1960s, metallicity has been considered as the main parameter governing the HB morphology. However the presence of clusters with the same [Fe/H] but different HB, suggests that at least another parameter is needed (see Catelan\,2009; Dotter et al.\,2010; Gratton et al.\,2010; Milone et al.\,2014 and references therein for recent review). 
For a long time helium has been considered a viable second-parameter candidate (D'Antona et al.\,2002; D'Antona \& Caloi\,2004).
This idea was however difficult to test as it requires the determination of helium in a large number of GCs and these measurements were challenging. 

There are three main techniques used in literature to estimate the helium abundance of stellar populations in GCs. 
Firstly, helium can be inferred from spectroscopy of 
very warm stars, like HB stars (see Villanova et al.\,2009, 2012; Marino et al.\,2014). 
If this method 
is able to provide direct determination of helium, it can be applied to stars in a limited temperature interval where spectroscopic analysis is difficult.  Indeed, highly excited He lines are not visible in stars cooler than temperatures $\sim$8500~K, 
while stars 
with temperatures higher than $\sim$11500~K 
 are subjected to phenomena like He sedimentation and the levitation of metals which alter the original surface abundance. 
Furthermore, 
since the distinct stellar populations in GCs usually occupy different regions along the HB (Marino et al.\,2011, 2013; Lovisi et al.\,2012; Gratton et al.\,2011, 2012, 2013), the HB segment with T$\sim$8500-11500~K provides partial information only.
A second spectroscopic approach is based on 
the analysis 
of chromospheric infrared lines in red-giant stars (Pasquini et al.\,2011; Dupree et al.\,2014), as done so far for a
few stars in NGC\,2808 and $\omega$\,Centauri only. However, chromospheric lines are very difficult to shape and sophisticated choromospheric models are required to infer stellar abundances.

A third approach to infer helium variations in GCs is  
 based photometry. Indeed, recent work suggests that
the analysis of multiple MSs and RGBs in the color-magnitude diagrams (CMDs) may provide a more efficient method to infer the helium content of multiple stellar populations in GCs (see Milone et al.\,2013 and references therein). 
In visual bands, helium impacts the luminosity of the MS star through the stellar temperature (D'Antona et al.\,2002, 2005). Since helium-enhanced stars have bluer colors than helium-normal stars with the same luminosity, multiple MSs can provide useful tool to infer the helium abundance of stellar populations through their color separation. 

Recent investigations on multiple sequences based on high precision photometry suggest that multiple stellar populations with different helium content could be a common feature of GCs (Milone et al.\,2014).
 The helium variation measured so far ranges from $\Delta$Y$\sim$0.01 in the cases of NGC\,6397 and NGC\,288 (Milone et al.\,2012b; Piotto et al.\,2013) up to $\sim$0.12 or more in NGC\,2808 and $\omega$\,Centauri (Piotto et al.\,2007; Bedin et al.\,2004; Norris\,2004; King et al.\,2012). 
 The internal helium variation in GCs seems to correlate with both the cluster mass and the color extension of the HB: massive clusters and GCs with very-extended HBs also host stellar populations that are highly helium enhanced (Milone et al.\,2014). However, I note that these conclusions are based on a small sample of GCs, which includes those objects where high-precision photometry was available to allow the helium measurements 
from multiple sequences.

The massive NGC\,6266 is an ideal candidate to investigate the relation between cluster mass, HB morphology and the helium content of multiple stellar populations. This GC hosts two main stellar populations with different content of C, Mg, Al, and Na (Yong et al.\,2014) and is among the ten most luminous Galactic GCs ($M_{\rm V}$=$-$9.18, Harris\,1996 updated as in 2010). It exhibits a very extended HB that is well populated both on the red and the blue side of the RR\,Lyrae instability strip (e.g.\,Piotto et al.\,2002; Beccari et al.\,2006; Valenti et al.\,2007), and hosts a large populations of RR\,Lyrae stars  whose periods 
agree with the Oosterhoff type I group 
(Contreras et al.\,2005, 2010). 
In this paper, I will exploit {\it Hubble Space Telescope} ({\it HST}) photometry to investigate, for the first time, multiple populations along the MS of NGC\,6266 and estimate their helium content. 
\section{Data and data analysis}
To investigate multiple stellar populations in NGC\,6266 I have used images collected with the Wide-Field Channel of the Advanced Camera for Surveys (WFC/ACS) and the Ultraviolet and Visual channel of the Wide Field Camera 3 (UVIS/WFC3) on board of {\it HST}. Details on the dataset are provided in Table~1.
The poor charge-transfer efficiency (CTE) in the WFC/ACS and UVIS/WFC3 images have been corrected by using the recipe and the software by Anderson \& Bedin\,(2010).
\begin{table*}
\caption{Description of the data set used in this paper.}
\begin{tabular}{cccccc}
\hline\hline
INSTR. & DATE & N$\times$EXPTIME & FILTER & PROGRAM & PI \\
\hline
WFC3/UVIS & Sep 18 2010 & 4$\times$35s+5$\times$393s+5$\times$421s & F390W & 11609 & J.\,Chaname \\
 ACS/WFC  & Aug 01 2004 & 200s+2$\times$340s & F435W & 10120 & S.\,Anderson\\
 ACS/WFC  & Aug 01 2004 & 30s+120s+3$\times$340s & F625W & 10120 & S.\,Anderson \\
 ACS/WFC  & Aug 01 2004 & 340s+3$\times$350s+3$\times$365s+3$\times$375s & F658N & 10120 & S.\,Anderson \\
\hline\hline
\end{tabular}\\
\end{table*}
Photometry and astrometry has been performed with {\it kitchen\_sync2}, which is a software program developed by Jay Anderson and mostly based on {\it kitchen\_sync} (Anderson et al.\,2008).
It uses different methods to measure bright and faint stars. Briefly, astrometric and photometric measurements of bright stars have been performed in each image, independently, by using the best point-spread function (PSF) model available, and later combined. To do this, I used library models from Anderson \& King (2006) and Anderson et al.\,(in preparation) and accounted for small focus variations due to the `breathing' of {\it HST}, by deriving for each exposure a spatially-constant perturbation. 
Furthermore, the flux and the position can also be determined by fitting for each star simultaneously all the pixels in all the exposures. This approach works better for very faint stars, which can not be robustly measured in every individual exposure. Stellar positions have been corrected for geometrical distortion by using the solution provided by Anderson \& King\,(2006) and Bellini, Anderson \& Bedin\,(2011). I refer the reader to the paper by Anderson et al.\,(2008) for further details.

The software by Anderson et al.\,(2008) provides a number of indexes that can be used as diagnostics of the quality of photometry. Since the study of multiple stellar population in GCs requires high-accuracy photometry, I have used these indexes to select a sub-sample of stars that have small astrometric errors, are relatively isolated, and well fitted by the PSF as in Milone et al.\,(2009, Sect.\,2.1). The study of NGC\,6266 is limited to this high-precision sub-sample of stars, which consists of about the 84\% of the total number of measured sources.

\begin{centering}
\begin{figure*}
 \includegraphics[width=14.5cm]{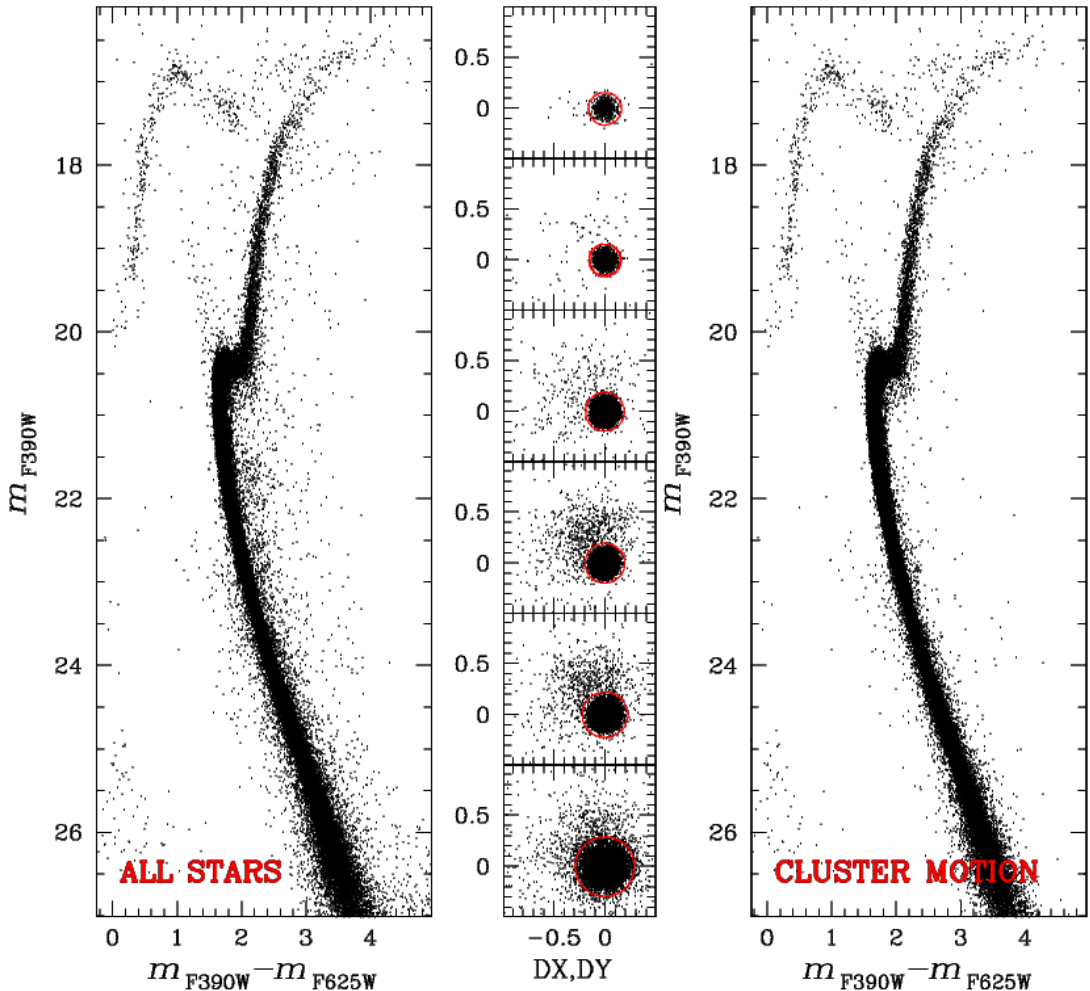}
 \caption{ \textit{Left:} $m_{\rm F390W}$ vs.\,$m_{\rm F390W}-m_{\rm F625W}$ CMD of all the stars in the field of view of NGC\,6266. \textit{Middle:} vector-point diagram of proper motions in ACS/WFC-pixel units for stars in six intervals of F390W magnitude. Stars within the red circles are considered probable cluster members. \textit{Right:} CMD for probable cluster members only.}
 \label{fig:PMs}
\end{figure*}
\end{centering}

The photometry has been calibrated by following the recipe by Bedin et al.\,(2005) and the zero points provided by the STScI web page for WFC/ACS and WFC3/UVIS\footnote{http://www.stsci.edu/hst/wfc3/phot\_zp\_lbn, http://www.stsci.edu/hst/acs/analysis/zeropoints/zpt.py}.
The $m_{\rm F390W}$ vs.\,$m_{\rm F390W}-m_{\rm F625W}$ CMD of all the stars in the field of view that pass the criteria of selection is shown in the left panel of Fig.~\ref{fig:PMs}. 

Since NGC\,6266 is projected on a rich Galactic field in the direction of the Galactic bulge, separating cluster members from foreground/background objects is an important step towards an appropriate study of the cluster's CMD. 
To select a sample of {\it bona-fide} cluster members I have derived proper motions by  comparing the stellar positions measured from 2004 images and the positions derived from data collected in 2010 and following the procedure described in detail by Anderson \& King (2004), Anderson \& van der Marel\,(2010), and Piotto et al.\,(2012). 

The middle panels of Fig.~\ref{fig:PMs} show the vector-point diagram of proper motions in ACS/WFC pixel, over the 6.1-year baseline spanned by the data, for stars in six magnitude intervals. 
 Since the zero point of the motion corresponds to the average motion of cluster stars the bulk of stars around is mostly populated by cluster members. The stars with clearly different motions are likely field objects and I used the red circles to select a sample of stars with cluster-like proper motion. The radius of each circle has been chosen by eye with the criterion of rejecting the most evident field stars. The CMD of stars with cluster motion is plotted in the right panel of Fig.~\ref{fig:PMs}.  

\subsection{Differential Reddening}
\label{sec:DR}
NGC\,6266 suffers for significant amounts of extinction (E(B$-$V)=0.47, Harris\,1996, updated as in 2010) and large spatial variations of reddening (Alonso-Garc{\'{\i}}a et al.\,2011).
Figure~\ref{fig:PMs} reveals that the CMD of the cluster exhibits broad RGB, HB, and SGB.  Since the photometric color error is smaller than $\sim$0.02 mag for these bright stars, one can expect that most of the broadening is caused by variations of reddening across the field of view. In addition to differential reddening, small unmodelable PSF variations can introduce small shifts in the photometric zero point which depend on the star location in the chip and result in a non-intrinsic broadening of the sequences in the CMD (Anderson et al.\,2008).
Previous works show that these PSF variations affect each filter in a different way and produce a small color variation which is typically $\sim$0.005~mag (Anderson et al.\,2008,2009; Milone et al.\,2012c).
 
 Correcting the photometry for both differential reddening and PSF-related variations of the photometric zero-point is a crucial step to investigate multiple stellar populations in NGC\,6266. To this aim I used a method mostly based on the recipe from Milone et al.\,(2012c, Sect.~3). 

Briefly, I have first determined a map of differential reddening across the field of view of NGC\,6266 and used it to correct the photometry.  To do this I started to define the fiducial line of cluster stars along the MS, the SGB, and the RGB. Then I have selected a sample of cluster members with high photometric quality  as reference stars and calculated for each of them the distance from the fiducial along the reddening line.
 To determine the direction of the reddening vector, I used the extinction rates of ACS/WFC bands provided by Sirianni et al.\,(2005) for a G2 star: A(F435W)=4.081\,E(B$-$V), A(F625W)=2.637\,E(B$-$V), A(F658N)=2.525\,E(B$-$V), and assumed A(F390W)=4.573\,E(B$-$V) (Aaron Dotter, private communication).    
 Only stars along the SGB, the bright MS, and the faint RGB have been chosen as reference stars because the angle between the reddening vector and the fiducial line is closer to 90$^{\rm o}$ and the color and magnitude displacement due to differential reddening can be properly separated from the random shift due photometric errors.
I assumed as the differential reddening of each star the median distance of the nearest 75 reference stars, while the corresponding uncertainty has been estimated as the 68.27$^{\rm th}$ percentile of the 75 distances divided by the square root of 74. The reference star has been excluded in the determination of its own differential reddening. The values of reddening have been converted into absorption in the WFC/ACS F435W, F625W, and F658N bands and in the WFC3/UVIS F390W band by using the relations above. Photometry in each band has been corrected for differential reddening by subtracting to the magnitude of each star the corresponding absorption.
I refer to the paper by Milone et al.\,(2012c) for further details on the differential reddening correction.
  
\begin{centering}
\begin{figure*}
 \includegraphics[width=14.5cm]{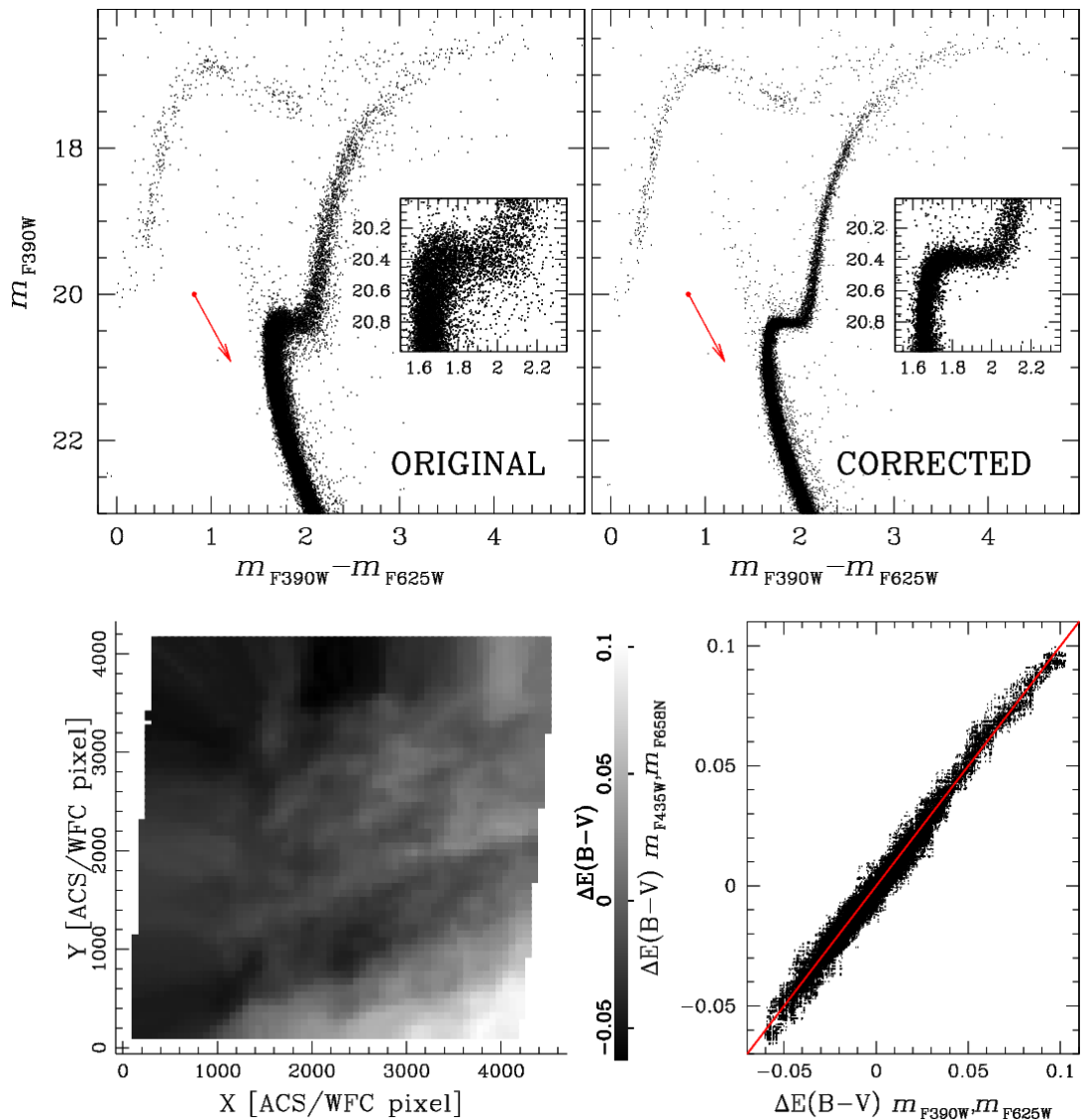}
 \caption{ 
\textit{Upper Panels:} Comparison of the original $m_{\rm F390W}$ vs.\,$m_{\rm F390W}-m_{\rm F625W}$ CMD of NGC\,6266 (left) and the CMD corrected for differential reddening (right). A zoom of each CMD around the SGB is plotted in the insets. The reddening vectors are indicated with red arrows.
\textit{Lower Panels:} Map of the average differential reddening (left). The levels of gray correspond to different E(B$-$V) values as indicated by the scale on the right. The right panel compares of the values of differential reddening inferred from the $m_{\rm F625W}$ vs.\,$m_{\rm F390W}-m_{\rm F625W}$ and the $m_{\rm F658N}$ vs\,$m_{\rm F435W}-m_{\rm F658N}$ CMD. The red line indicates the perfect agreement. 
}
 \label{fig:red}
\end{figure*}
\end{centering}

As an example, the upper panels of Fig.~\ref{fig:red} compare the original $m_{\rm F390W}$ vs.\,$m_{\rm F390W}-m_{\rm F625W}$ CMD of NGC\,6266 (left) and the CMD corrected for differential reddening (right). The insets show the region around the SGB, where the effect of differential reddening is more evident.
An inspection of the obtained differential reddening map, 
in the lower-left panel of Fig.~\ref{fig:red}, reveals significant spatial variations, up to $\sim$0.15 mag in E(B$-$V), across the field of view. The error on differential-reddening determination, is typically $\sigma_{\rm E(B-V)} \sim$0.003 mag and never exceeds 0.006 mag.

This procedure has been applied to the each CMD analyzed in this paper, separately. 
To investigate if the obtained differential reddening depends on the adopted color and magnitude, I compare results from different CMDs.
 With photometry in four bands, I can generate two independent CMDs: $m_{\rm F625W}$ vs.\,$m_{\rm F390W}-m_{\rm F625W}$ and $m_{\rm F658N}$ vs.\,$m_{\rm F435W}-m_{\rm F658N}$. The values of differential reddening obtained from each of them, $\Delta$E(B$-$V)$_{\rm F390W,F625W}$ and $\Delta$E(B$-$V)$_{\rm F435W,F658N}$, are compared in the lower-right panel of Fig.~\ref{fig:red}. They differ on average by $\Delta$E(B$-$V)=0.002, with a random median scatter of 0.006~mag which is due, in part, to PSF-related variations of the photometric zero-point along the chip. Difference are smaller when I compare $\Delta$E(B$-$V)$_{\rm F390W,F625W}$ with results from the other CMDs used in this paper.

\subsection{Artificial Stars}
 The artificial-star (AS) experiments have been performed by using the recipe and the software described in detail by Anderson et al.\,(2008). 
 Briefly, a list of 10$^{5}$ stars has been generated and placed 
along the fiducial line of the MS, the SGB, and the RGB of NGC\,6266. The list includes the coordinates of the stars in the reference frame and the magnitudes in F390W, F435W, F625W, and F658N bands. ASs have been distributed across the field  of view according to the overall cluster distribution as in Milone et al.\,(2009).

For each star in the input list, the software by Anderson et al.\,(2008) adds, in each image, a star with appropriate flux and position and measures it by using the same procedure as for real stars. I considered an AS as found when the input and the output position differ by less then 0.5 pixel and the input and the output flux by less than 0.75 mag.

The software provides for ASs the same diagnostics of the photometric quality as for real star. I applied to ASs the same procedure used for real stars to select a sub-sample of relatively-isolated stars with small astrometric errors, and well fitted by the PSF.
ASs have been used to estimate errors of the photometry used in this paper.

\section{Multiple populations along the MS}
\label{sec:CMD}
\begin{centering}
\begin{figure*}
 \includegraphics[width=14.5cm]{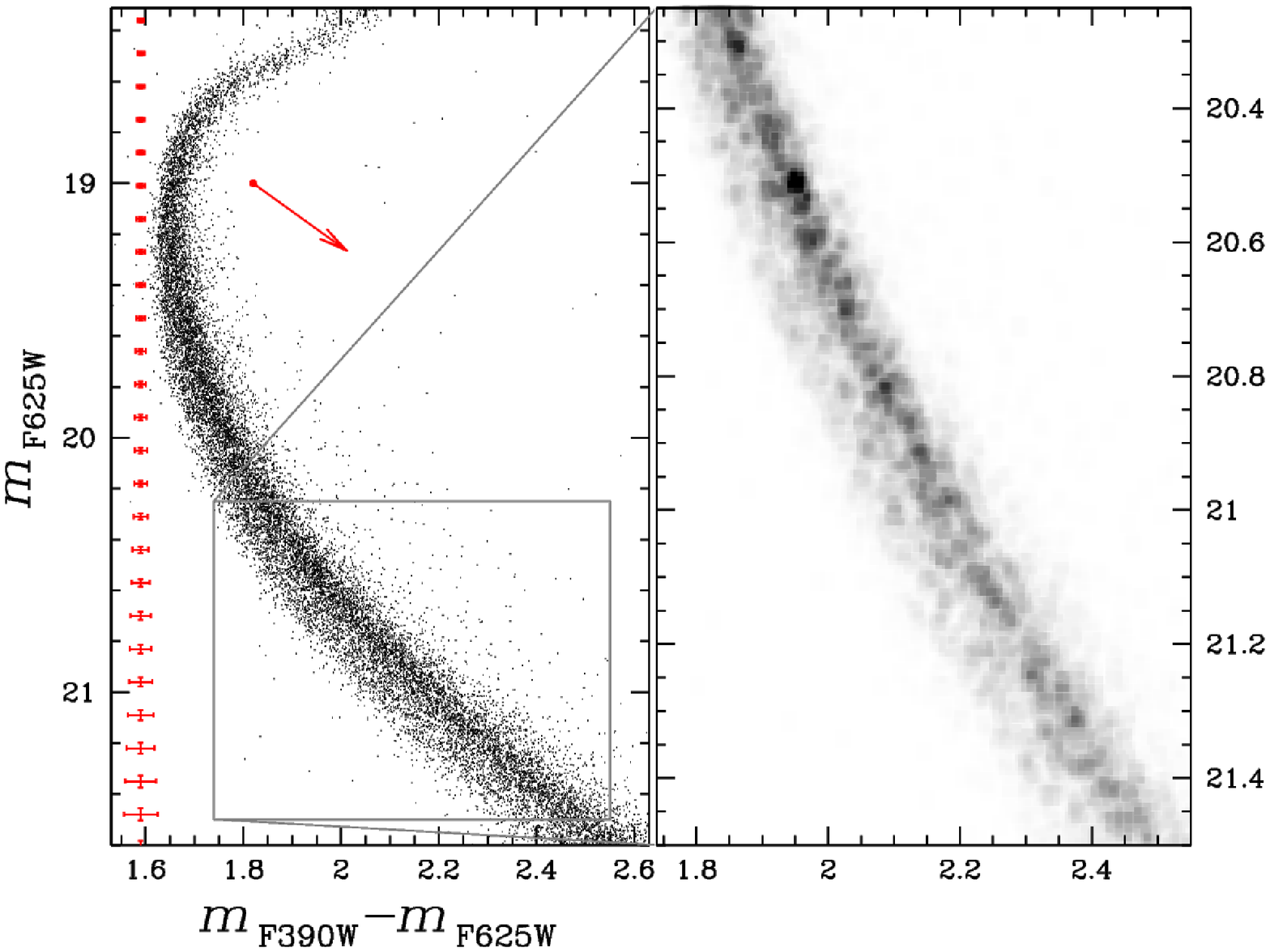}
 \caption{$m_{\rm F625W}$ vs.\,$m_{\rm F390W}-m_{\rm F625W}$ CMD for MS and SGB stars in NGC\,6266. The arrow indicates the reddening direction. The right panel shows the Hess diagram of MS stars with $20.25<m_{\rm F625W}<21.5$.}
 \label{fig:fig2}
\end{figure*}
\end{centering}

The left panel of Fig.~\ref{fig:fig2} shows a zoom of the $m_{\rm F625W}$ vs.\,$m_{\rm F390W}-m_{\rm F625W}$ CMD of NGC\,6266 around the MS and the SGB. 
I have applied the corrections for differential reddening and for PSF-related color variations, as described in Sect.~\ref{sec:DR}, and have excluded stars with radial distance smaller than 50 arcsec from the cluster center, to avoid the most crowded regions.
 The observational errors are represented with red error bars on the left of the CMD and account for both photometric errors, as derived from AS experiments, and uncertainties on the differential-reddening correction.

The MS of NGC\,6266 is broad with some hint of bimodality. The majority of MS stars defines a red MS but there is a tail of stars on the blue side of the main MS. The MS broadening is more evident in the magnitude interval $20.25<m_{\rm F625W}<21.5$, as better visualized by the Hess diagram in the right panel of Fig.~\ref{fig:fig2}.
  The observed color spread for MS stars is much larger than that expected from  observational error, which, in this magnitude interval, ranges from $\sim$0.015 to $\sim$0.035~mag.
I note that this observed multiple MS in NGC\,6266 can not be the consequence of residual differential reddening for two reasons:
{\it (i)} the red and the blue MS can be detected in any region of the field of view;
{\it (ii)} the effect of differential reddening would be more evident at the level of the SGB and the MS turn off, where, as indicated by the arrow in Fig.~\ref{fig:fig2}, the reddening vector is almost perpendicular to the sequences. Instead, NGC\,6266 exhibits a narrow color distribution in this region of the CMD.

\begin{centering}
\begin{figure}
 \includegraphics[width=9.0cm]{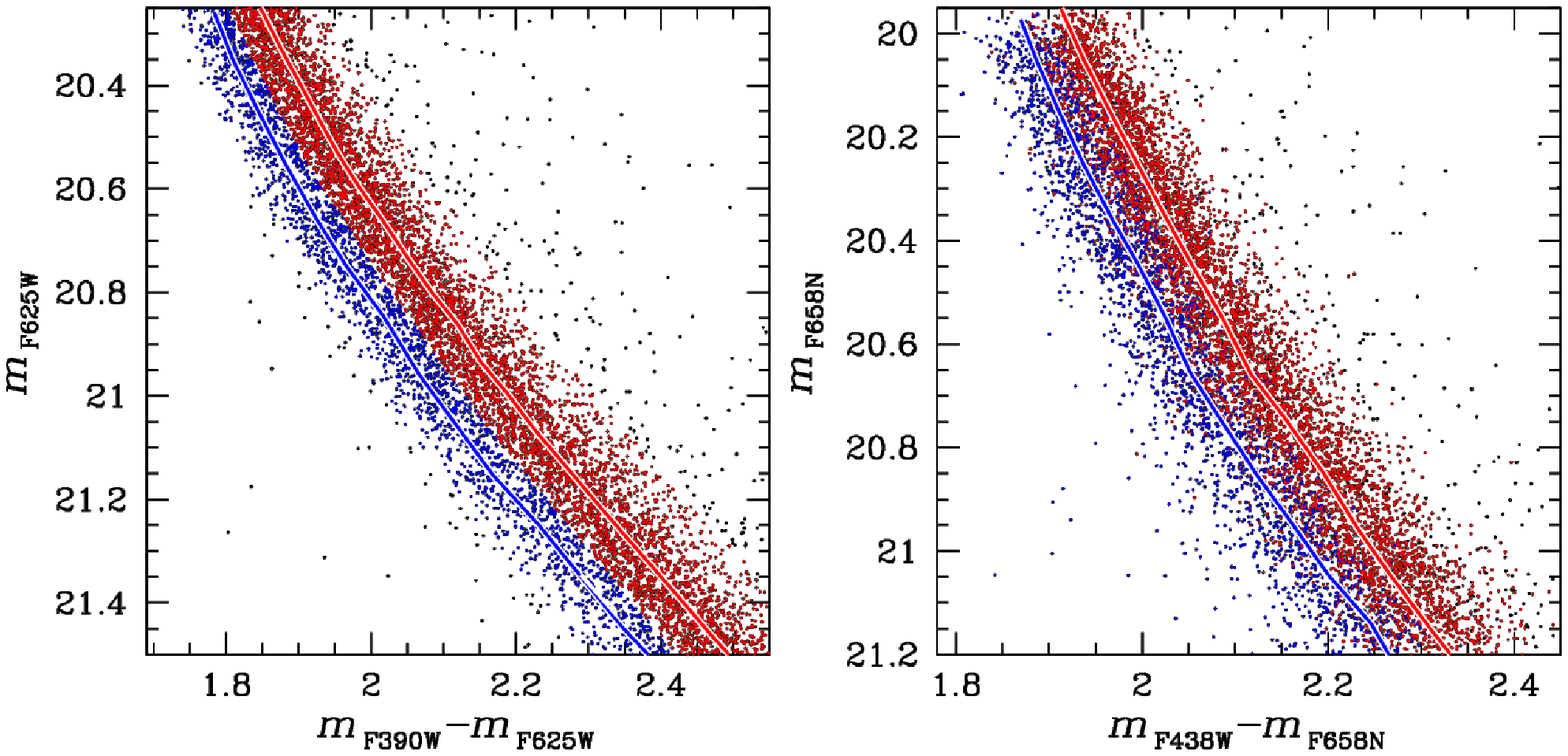}
 \caption{ $m_{\rm F625W}$ vs. $m_{\rm F390W}-m_{\rm F625W}$ (left) and $m_{\rm F658N}$ vs. $m_{\rm F435W}-m_{\rm F658N}$ CMD of MS stars in NGC\,6266 (right). The sample of red-MS and the blue-MS stars, defined in the left-hand CMD are colored red and blue, respectively. The fiducial lines of the red and blue MS are superimposed on each CMD.}
 \label{fig:SEPraw}
\end{figure}
\end{centering}

To further investigate whether the MS broadening is entirely due to photometric errors or it is also intrinsic, I have adapted to NGC\,6266 the method introduced by Anderson et al.\,(2009) in their study of multiple populations along the MS of 47\,Tuc.
Briefly, I have used the two CMDs shown in Fig.~\ref{fig:SEPraw} that are derived from two independent datasets and 
identified by eye the two groups of red and blue MS stars in the $m_{\rm F625W}$ vs.\,$m_{\rm F390W}-m_{\rm F625W}$ CMD.
(represented in red and blue, respectively, in both CMDs). 
Then, I have determined the fiducial line of red- and blue-MS in each CMD
by calculating the median colors and magnitude in successive short intervals of magnitude and interpolated these median points with a cubic spline.

As discussed by Anderson and collaborators, if the MS broadening is only due to errors, a star that is red or blue in the $m_{\rm F625W}$ vs. $m_{\rm F390W}-m_{\rm F625W}$ CMD has the same probability of being either red or blue in the $m_{\rm F658N}$ vs. $m_{\rm F435W}-m_{\rm F658N}$ ones. 
 On the contrary, the fact that the fiducial of the red and the blue MS are clearly separated in both CMDs demonstrates that the color of each star is maintained quite well, 
with only a small scatter due to photometric errors. 
This behaviour is a clear mark of an intrinsic color spread. 

\begin{centering}
\begin{figure*}
 \includegraphics[width=14.5cm]{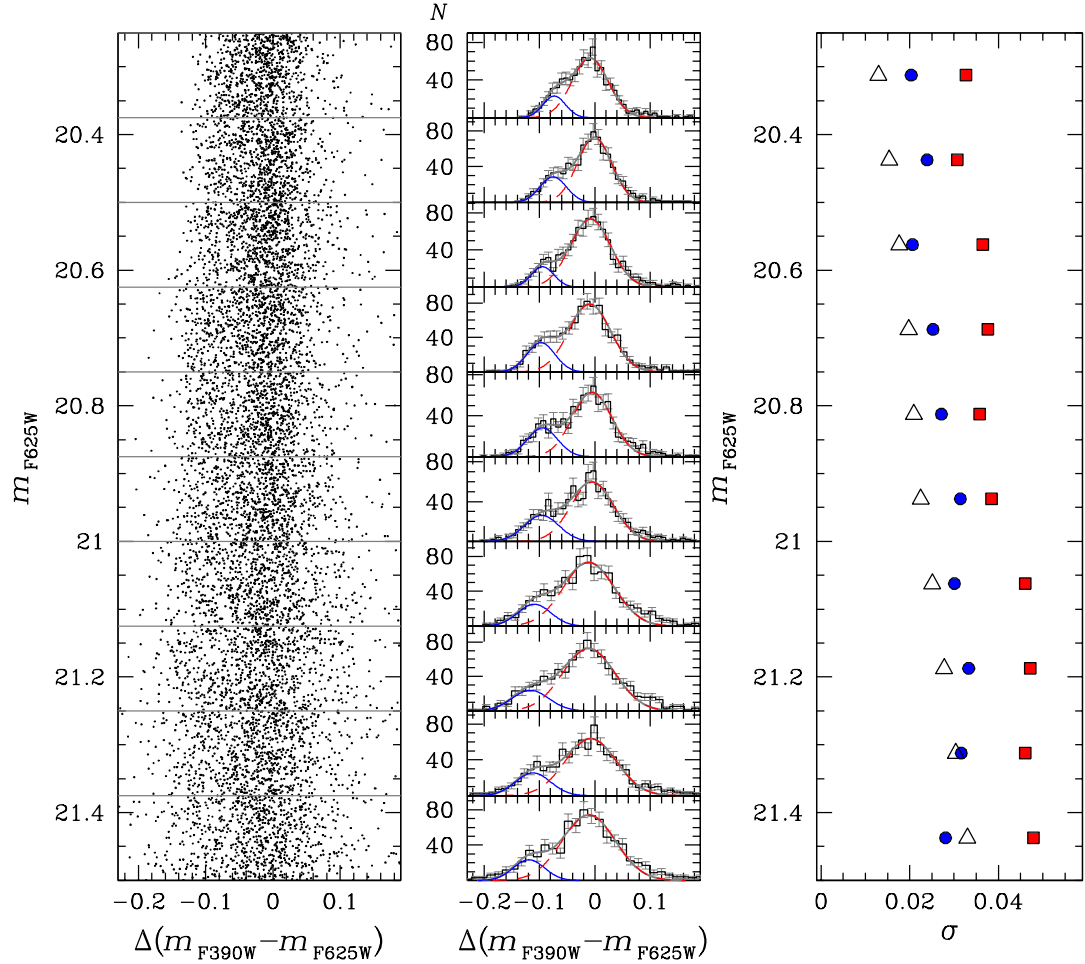}
 \caption{\textit{Left:} Verticalized $m_{\rm F625W}$ vs. $\Delta$($m_{\rm F390W}-m_{\rm F625W}$) diagram for MS stars with $20.25<m_{\rm F625W}<21.5$. 
 \textit{Middle:} Histogram distribution of $\Delta$($m_{\rm F390W}-m_{\rm F625W}$) for stars in the ten intervals of F625W magnitude marked by gray lines in the left panel. The gray lines are the best-fitting least-squares bi-Gaussian functions, whose components are represented with blue-continuous and red-dashed lines. 
\textit{Right:} $m_{\rm F625W}$ as a function of the observed $\Delta$($m_{\rm F390W}-m_{\rm F625W}$) dispersion for the red MS (squares), the blue MS (circles), and the dispersion due to observational errors (triangles).}
 \label{fig:Pratio}
\end{figure*}
\end{centering}

To determine the fraction of red-MS and blue-MS stars, I used the $m_{\rm F625W}$ vs. $m_{\rm F390W}-m_{\rm F625W}$ CMD plotted in the left panel of Fig.~\ref{fig:SEPraw} and applied the procedure illustrated in Fig.~\ref{fig:Pratio}. The verticalized $m_{\rm F625W}$ vs. $\Delta$($m_{\rm F390W}-m_{\rm F625W}$) diagram 
(left panel of Fig.~\ref{fig:Pratio}) has been obtained by subtracting from the $m_{\rm F390W}-m_{\rm F625W}$ color of each star, the color of the red fiducial line at the corresponding  F625W magnitude.
 This diagram has been divided into ten bins of 0.125 magnitude as indicated by the gray horizontal lines. The histogram distribution of $\Delta$($m_{\rm F390W}-m_{\rm F625W}$) for stars in each bin has been plotted in the middle panel of Fig.~\ref{fig:Pratio} and 
fitted by a bi-Gaussian function by using the least-squares method. The fraction of red-MS and blue-MS stars has been derived, for each magnitude bin, from the area of the two Gaussians. From the average of the ten independent measurements,
the red and the blue MS result to host the 79$\pm$1\% and the 21$\pm$1\% of 
 stars, respectively. The errors are derived as the random mean scatter of the ten measurements divided by the square root of nine.
 
The F625W magnitude has been plotted as a function of the dispersion obtained for the red MS ($\sigma_{\rm red-MS}$, squares), the blue MS ($\sigma_{\rm blue-MS}$, circles) from the best-fit Gaussians, and the dispersion due to observational errors ($\sigma_{\rm err}$, triangles) in the right panel of Fig.~\ref{fig:Pratio}. 
While the color width of blue-MS stars is comparable to the spread due to observational errors\footnote{ The fact that the broadening blue MS is slightly larger than expectation from observational errors is due to a limitation of AS test. Indeed while ASs are measured with the same PSF used to generate them, the PSF model for a real star can slightly differ from the real stellar profile. As a consequence, measurements of real stars will be affected by the errors in the PSF model, but AS measurements will not (see Anderson et al.\,2008; Milone et al.\,2009 for details). However, the additional dispersion do not allow to exclude a small intrinsic spread.} 
 the color width of the red MS is significantly larger than that of the blue MS over the whole range of analyzed magnitude.
This additional spread, 
 which can be quantified as $\sigma_{\rm red-MS}^{\rm int}=\sqrt{\sigma_{\rm red-MS}^{2}-\sigma_{\rm err}^{2}}$,
 suggests that the red MS exhibits an intrinsic color broadening.
 The values of $\sigma_{\rm blue-MS}$, $\sigma_{\rm red-MS}$, $\sigma_{\rm err}$, and $\sigma_{\rm red-MS}^{\rm int}$, are listed in Table~2 for different intervals of F625W magnitude.

\begin{table}
\scriptsize
\caption{ Color dispersion observed for blue-MS stars, red-MS stars and dispersion due to observational errors in ten magnitude intervals. The intrinsic color dispersion of the red MS is listed at the last column.  \label{tab:sigma}}
\centering
\begin{tabular}{ccccc}

\hline
\hline
       $m_{\rm F625W}$ &   $\sigma_{\rm blue-MS}$  &  $\sigma_{\rm red-MS}$ &  $\sigma_{\rm err}$ & $\sigma_{\rm red-MS}^{\rm int}$     \\
\hline
       20.25 - 20.38 &       0.020 &       0.032 &       0.013 &        0.029 \\ 
       20.38 - 20.50 &       0.024 &       0.031 &       0.015 &        0.027 \\ 
       20.50 - 20.62 &       0.021 &       0.037 &       0.018 &        0.032 \\ 
       20.62 - 20.75 &       0.025 &       0.038 &       0.020 &        0.032 \\ 
       20.75 - 20.88 &       0.027 &       0.036 &       0.021 &        0.029 \\ 
       20.88 - 21.00 &       0.031 &       0.038 &       0.022 &        0.031 \\ 
       21.00 - 21.12 &       0.030 &       0.046 &       0.025 &        0.039 \\ 
       21.12 - 21.25 &       0.033 &       0.047 &       0.028 &        0.038 \\ 
       21.25 - 21.38 &       0.032 &       0.046 &       0.030 &        0.035 \\ 
       21.38 - 21.50 &       0.028 &       0.048 &       0.033 &        0.035 \\ 
\hline
\hline
\end{tabular}
\end{table}

A spread or split of the MS about one magnitude below the MS turn off could be due either to binaries or to multiple stellar populations.
Binary systems composed by two MS stars (MS-MS binaries) are seen in clusters as a spread sequence on the red of the MS. However, to ascribe the red-MS stars to a sequence of binaries would require that about 88\% of the MS stars in NGC\,6266 are in binary systems, in contrast with observations of massive GC, where the fraction of MS-MS binaries is smaller than a few percent (Milone et al.\,2012c). Besides, to reproduce the narrow color distribution of red-MS stars one could make the outlandish hypothesis that most binaries have a primary/secondary mass ratio in the narrow interval $\sim$0.7-0.8. I conclude, that very unlikely binaries can explain the observed MS spread.

 A more-plausible hypothesis is that the red and the blue MS of NGC\,6266 correspond to distinct stellar populations, in close analogy with what observed in most GCs. This is supported by results presented in the next Section, where I demonstrate that observations are consistent with two populations with different content of helium and light elements.

\section{The helium content of the stellar populations}
\label{sec:elio}
\begin{centering}
\begin{figure*}
 \includegraphics[width=12.5cm]{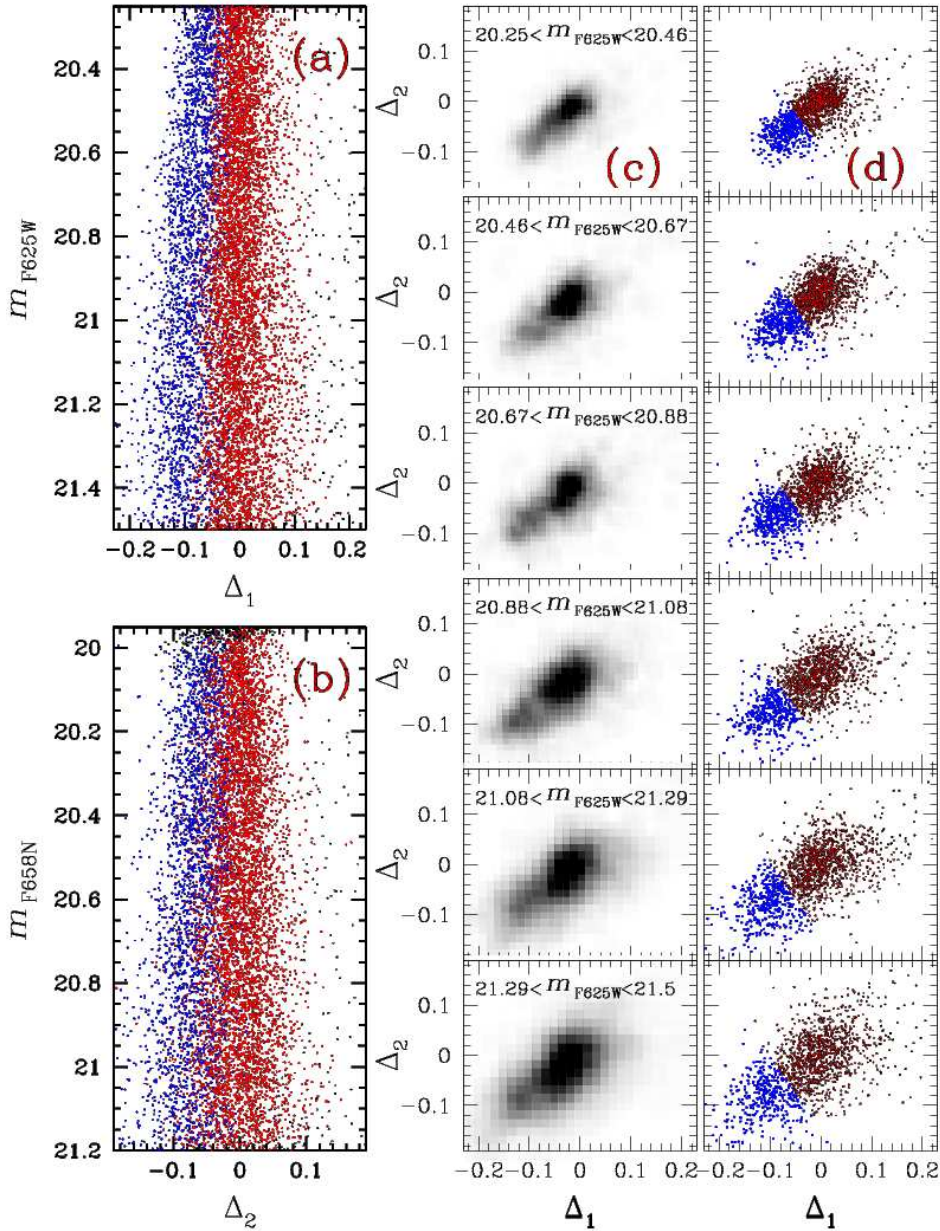}
 \caption{Verticalized $m_{\rm F625W}$ against $\Delta_{1}$ and $m_{\rm F658N}$ against $\Delta_{2}$ diagram for MS stars (panels a and b). The Hess-diagrams of $\Delta_{2}$ vs. $\Delta_{1}$ for stars in six intervals of $m_{\rm F625W}$ are shown in panels c. Red and blue stars in panels a, b, and d, represent the two groups of red-MS and blue-MS stars, defined in the $\Delta_{2}$ vs. $\Delta_{1}$ diagrams of panel (d).}
 \label{fig:sep}
\end{figure*}
\end{centering}

To further investigate multiple MSs, in Fig.~\ref{fig:sep} I show the verticalized $m_{\rm F625W}$ vs. $\Delta_{1}$ (panel a) and $m_{\rm F658N}$  vs. $\Delta_{2}$ diagram (panel b) for MS stars obtained from two independent dataset.
 The value of $\Delta_{1}$ has been derived for each star, by subtracting from the $m_{\rm F390W}-m_{\rm F625W}$ color of that star, the corresponding color of the red fiducial line of Fig.~\ref{fig:SEPraw} at the same F625W magnitude. A similar approach has been used to determine $\Delta_{2}$.
In panels (c), I show the Hess diagram of $\Delta_{2}$ vs. $\Delta_{1}$ for MS stars in six F625W luminosity bins. The correlation between $\Delta_{2}$ and $\Delta_{1}$ confirms that the MS of NGC\,6266 exhibits an intrinsic color spread. The stellar distribution in each bin is clearly bimodal, with red-MS stars clustered around ($\Delta_{1}$:$\Delta_{2}$)$\sim$(0:0), plus a less-populated bump of stars with negative $\Delta_{1}$ and $\Delta_{2}$ associated with a blue MS.

 The $\Delta_{2}$ vs. $\Delta_{1}$ diagrams of Fig.~\ref{fig:sep} provide the opportunity to improve the previous identification of red-MS and blue-MS stars of Fig.~\ref{fig:SEPraw}, which was based on $m_{\rm F390W}$ and $m_{\rm F625W}$ only.
To this aim, I identify by eye two groups of red-MS and blue-MS stars, on the basis of their position in the $\Delta_{2}$ vs. $\Delta_{1}$ planes of panel (d) 
(coloured in red and blue, respectively, as also in panels a and b). 

\begin{centering}
\begin{figure*}
 \includegraphics[width=14.5cm]{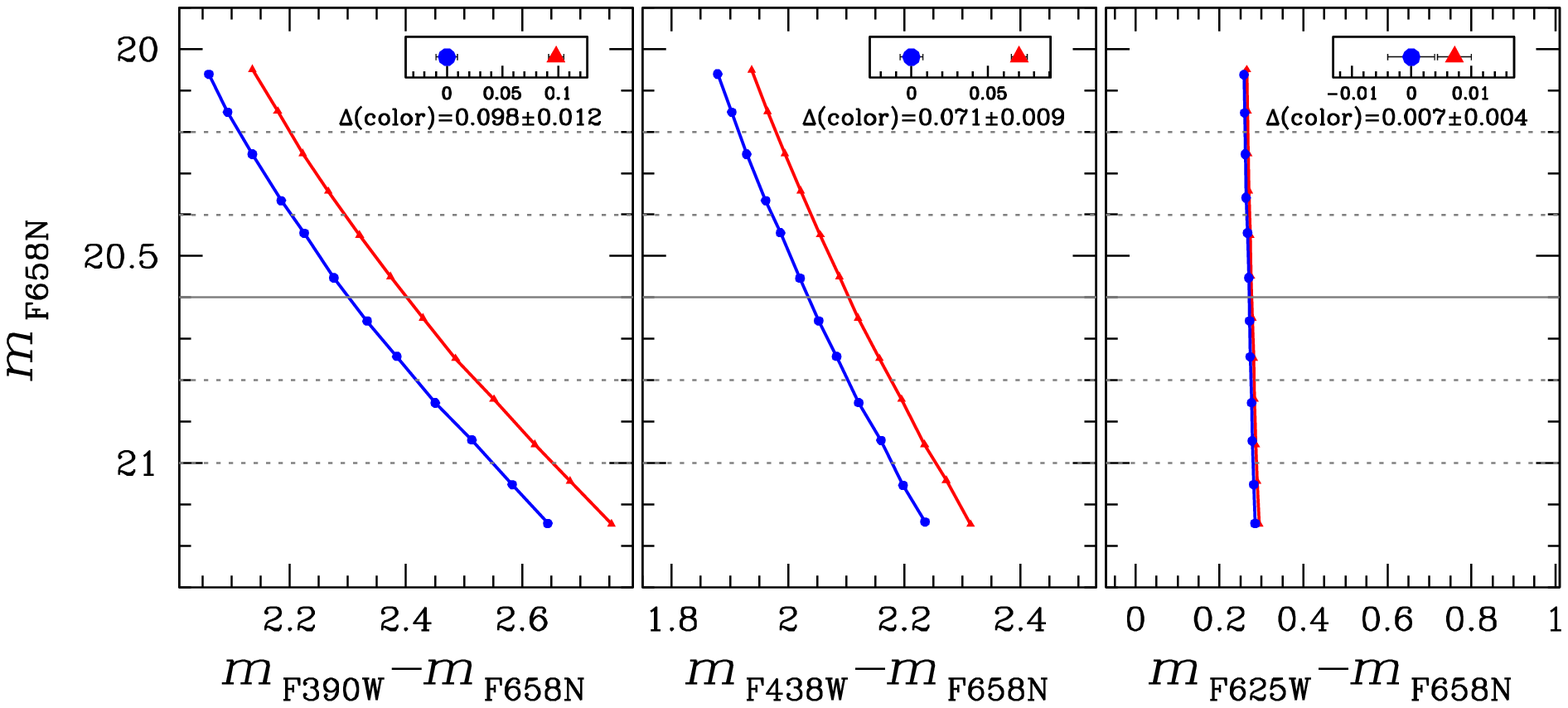}
 \caption{Fiducial lines of red and blue MS in the $m_{\rm F658N}$ vs.  $m_{\rm X}-m_{\rm F658N}$ CMD, where X= F390W (left), F435W (middle), and F625W (right). The $m_{\rm X}-m_{\rm F658N}$ color separation between red and blue MS quoted in each panel is estimated at $m_{\rm F658N}^{\rm CUT}$=20.6 (continuous line). The dotted-horizontal lines mark the other values of $m_{\rm F658N}^{\rm CUT}$.}
 \label{fig:fids}
\end{figure*}
\end{centering}

 To infer the average helium difference between the two MSs of NGC\,6266, I started to 
investigate the CMDs
$m_{\rm F658N}$ against 
$m_{\rm X}-m_{\rm F658N}$, where X= F390W, F435W, and F625W. For each CMD I have determined the fiducial line of the red and the blue MS, as represented in Fig.~\ref{fig:fids}, by using the method described in Sect.~\ref{sec:CMD}. 
Then, I have estimated the color difference between the red and blue fiducials ($\Delta$(color)) at the reference magnitudes $m_{\rm F658N}^{\rm CUT}$=20.2, 20.4, 20.6, 20.8, and 21.0~mag. 
Results, 
listed in Table~3, 
show that for each choice of $m_{\rm F658N}^{\rm CUT}$, the $m_{\rm F658N}$ vs. $m_{\rm F390W}-m_{\rm F658N}$ CMD provides the largest color difference between red and blue fiducial. The value of $\Delta$(color) decreases for shorter color baselines. 

Recent papers demonstrate that multiple MSs correspond to stellar populations with difference in helium (e.g.\,D'Antona et al.\,2002; Norris\,2004; Piotto et al.\,2005) and in light-elements abundance (Marino et al.\,2008; Sbordone et al.\,2011; Milone et al.\,2012a,b).
To infer the average helium difference between the two populations of NGC\,6266, theoretical spectra for red- and blue-MS stars have been calculated to 
compare the 
predicted colors with the observed ones. 
For the synthetic spectra, different
values of light-element abundances 
and different choices of helium have been considered for each population. Specifically, 
for red-MS stars primordial helium (Y=0.246) has been assumed, while 
for the blue MS a grid of model with helium ranging from Y=0.246 to 0.400 in steps of 0.001 has been generated.

To account for the chemical composition of stellar populations in NGC\,6266 I exploit results from the analysis of high-resolution spectra by Yong et al.\,(2014). They have analyzed seven bright giants and detected significant star-to-star variations in C, Mg, Al, O, and Na, that define the usual (anti)correlations 
observed in GCs, 
with the presence of two stellar groups: an O-rich (Na-poor) and O-poor (Na-rich) group of stars.
 Following Yong et al.\,(2014), I use for red- and blue-MS stars the average inferred abundances of O-rich and O-poor stars, respectively.
 Specifically, I assumed [C/Fe]=$-$0.4 and [O/Fe]=0.8 for red-MS stars, and [C/Fe]=$-$0.8 and [O/Fe]=0.2 for the blue MS. Since no nitrogen abundance are available for this cluster I arbitrarily assumed that the red and the blue MS have [N/Fe]=0.5 and [N/Fe]=1.6, respectively, as inferred by Bragaglia et al.\,(2010) for the red and the blue MS of NGC\,2808, which is another massive GC with an extended HB and almost the same metallicity as NGC\,6266.
 
For both MSs an uniform metallicity of [Fe/H]=$-$1.15 and alpha-enhancement [$\alpha$/Fe]=0.40 has been assumed, as derived from spectroscopy by Yong et al.\,(2014). 
I adopt reddening, E(B$-$V)=0.52, age of 12.5 Gyr, and a distance modulus, $(m-M)_{\rm V}$=15.55, that provide the best fit between isochrones from Dotter et al.\,2008 and the observed CMD.
Isochrones with primordial helium (Y=0.246) have been used to estimate effective temperature ($T_{\rm eff}$) and surface gravity (log~$g$) for MS stars with $m_{\rm F658N}$=$m_{\rm F658N}^{\rm CUT}$, as listed in Table~3; while
$T_{\rm eff}$ and log~$g$ values for helium-rich blue-MS stars have been inferred from helium-enhanced 
isochrones. 

\begin{table*}
\scriptsize
\caption{Color separation between blue and red MS and stellar parameters of the best-fitting model for red and blue MS for five values of $m_{\rm F658N}^{\rm CUT}$. The average helium difference is listed in the last column. \label{tab:parametri}}
\begin{tabular}{lcccccccc}
\hline\hline
 $m_{\rm F658N}^{\rm CUT}$ & $\Delta$($m_{\rm F390W}-m_{\rm F658N}$) & $\Delta$($m_{\rm F435W}-m_{\rm F658N}$) & $\Delta$($m_{\rm F625W}-m_{\rm F658N}$) & $T_{\rm eff}$ (red MS) & log\,$g$ (red MS) & $T_{\rm eff}$(blue MS) & log\,$g$ (blue MS)  & $\Delta$Y \\
\hline
    20.2 & 0.081$\pm$0.010 & 0.063$\pm$0.009 & 0.007$\pm$0.003 & 5673 & 4.54 & 5831 & 4.52 & 0.086 \\
    20.4 & 0.091$\pm$0.011 & 0.066$\pm$0.009 & 0.007$\pm$0.004 & 5548 & 4.57 & 5702 & 4.56 & 0.079 \\
    20.6 & 0.098$\pm$0.012 & 0.071$\pm$0.009 & 0.007$\pm$0.004 & 5414 & 4.60 & 5560 & 4.59 & 0.074 \\
    20.8 & 0.104$\pm$0.015 & 0.077$\pm$0.011 & 0.008$\pm$0.005 & 5276 & 4.62 & 5425 & 4.62 & 0.073 \\
    21.0 & 0.109$\pm$0.016 & 0.079$\pm$0.012 & 0.008$\pm$0.006 & 5136 & 4.64 & 5295 & 4.65 & 0.078 \\
\hline
\hline
\end{tabular}
\end{table*}

 Atmospheric models for the MS stars with $m_{\rm F658N}$=$m_{\rm F658N}^{\rm CUT}$ have been calculated by using the ATLAS12 code (Kurucz\,1993; Sbordone et al.\,2004) and have been used to generate synthetic spectra with SYNTHE (Kurucz \& Avrett\,1981) at a resolution of $R$=600 from 2800 to 10000\AA.

\begin{centering}
\begin{figure*}
 \includegraphics[width=8.5cm]{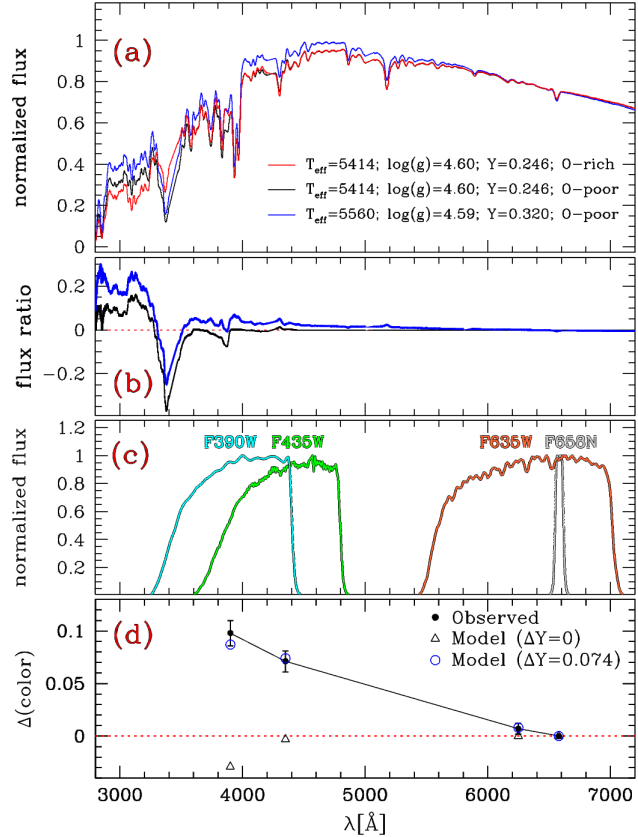}
 \caption{\textit{Panel a}: comparison of the synthetic spectra of O-rich star (red), and O-poor stars with Y=0.246 (black) and Y=0.320 (blue). The spectra correspond to MS star with $m_{\rm F658N}^{\rm CUT}$=20.6 (see text for details). \textit{Panel b}: Flux ratio between the spectra of the O-poor/Y-poor and the O-rich star (black) and the O-poor/Y-rich and the O-rich star (blue). \textit{Panel c}: Transmission curves of the {\it HST} filters used in this work. \textit{Panel d}: ${\it m}_{\rm X}-{\it m}_{\rm F658N}$ color separation between the fiducial lines of the two MSs at $m_{\rm F658N}^{\rm CUT}$=20.6 as a function of the central wavelength of the X filter, where X=F390W, F435W, F625W, and F658N. Observations are plotted with black filled circles, while the color difference derived from theoretical models are represented with open symbols.}
 \label{fig:spettri}
\end{figure*}
\end{centering}

Results are illustrated in Fig.~\ref{fig:spettri} for MS stars 
at $m_{\rm F658N}^{\rm CUT}$=20.6. Synthetic spectra for these stars are shown in panel (a), 
 which compares the spectrum corresponding to the O-rich composition (red), 
and those corresponding to the O-poor composition at Y=0.246 (black) and Y=0.320 (blue).
The flux ratios of the black and blue spectra with respect to the red ones are plotted in panel (b). 
Finally, panel (c) shows the normalized response of the UVIS/WFC3 and the ACS/WFC filters used in this paper.

Synthetic spectra of O-poor and O-rich stars have been integrated over the transmission curves of the F390W UVIS/WFC3 and the F435W, F625W, and F658N ACS/WFC filters to determine synthetic colors. The filled circles in the panel (d) of Fig.~\ref{fig:spettri} show the ${\it m}_{\rm X}-{\it m}_{\rm F658N}$ color difference  between the fiducial line of the two MSs at $m_{\rm F658N}^{\rm CUT}$=20.6 
 against the central wavelength of the given X filter; the color differences from theoretical models are represented with open symbols.
The O-poor model with primordial helium (triangles) 
provides a very poor fit with the observations
thus demonstrating that the two stellar populations of NGC\,6266 have different helium abundance.

To determine the value of Y that best matches the CMD sequences, I have compared observations with synthetic colors derived from helium-enhanced blue-MS stars 
with different values of helium assumed, 
ranging from Y=0.247 to 0.400 in steps of 0.001.
The best fit between observations at $m_{\rm F658N}^{\rm CUT}=$20.6 corresponds to the case where the O-poor population is enhanced in helium by $\Delta$Y=0.074~dex (blue circles). 

This procedure has been repeated for $m_{\rm F658N}^{\rm CUT}$=20.2, 20.4, 20.8, and 21.0~mag. The values of $T_{\rm eff}$, log($g$), and $\Delta$Y that provide the best fit are listed in Table~3 for each value of $m_{\rm F658N}^{\rm CUT}$. From the average I obtain $\Delta$Y=0.078$\pm$0.003, where the error is determined as the rms of the five determinations of $\Delta$Y divided by the square root of four. 

The helium abundance inferred from this method depends on the chemical composition assumed for the red and the blue MS. 
The main contribution may come from the nitrogen abundance, which affects the F390W magnitude trough the NH band around $\lambda \sim$3300 \AA.  To quantify the impact of the choice of N on the inferred $\Delta$Y, I repeated the same procedure above for two extreme scenarios where: i) the blue MS is highly N-enhanced by 2 dex with respect to the red MS and ii) the two MSs have the same N abundance. From the first scenario I infer a value of $\Delta$Y which is 0.011 dex larger than the previous estimate, while the later scenario corresponds to a $\Delta$Y lower than 0.007 dex with respect to the best estimate.

In order to investigate the impact of the adopted abundance of C and O on the estimate of helium abundance, I again repeated the procedure above by changing the values of [C/Fe] and [O/Fe] adopted for the blue MS by $\pm$0.4 dex. In all the cases the resulting $\Delta$Y is consistent with the best estimate within 0.001 dex. 
Similarly, I verified that variations in Al, Mg, and Na do not affect $\Delta$Y.

 It is worth noting that the adopted values of [C/Fe] come from spectroscopy of RGB stars. As pointed out by the referee, both theory and observations have shown that as a star ascend the RGB, carbon and nitrogen are exposed to mixing phenomena  which alter the original surface abundance of C and N, while maintaining constant C+N (e.g.\,Iben\,1967; Charbonnel\,1994; Gratton et al.\,2000; Martell et al.\,2008; Angelou et al.\,2011 and references therein).  To estimate the effect of mixing on the results of the paper, I assumed that both red- and blue-MS stars have 0.3-dex higher C abundance lower than that inferred from the RGB by Yong et al.\,(2014). The adopted carbon variation between MS and RGB stars approximately matches the predictions from models by Angelou et al.\,(2011) for CN-weak and CN-strong stars in M\,3. [N/Fe] has been determined with the criteria that the sum of carbon plus nitrogen remains constant.
The obtained nitrogen abundance of red-MS stars is 0.3-dex lower than that inferred from the RGB, while, in the case of the helium-rich population of NGC\,6266, 
 an enhancement in C by 0.3 dex leaves N almost unchanged.
I again derived $\Delta$Y by using the procedure above but assuming [C/Fe]=$-$0.1 and [N/Fe]=0.2 for the MS, and  [C/Fe]=$-$0.5 and [N/Fe]=1.6. I infer a value of $\Delta$Y which is 0.004 dex larger than the best estimate.

This investigation shows that, while the derived helium abundance does not depend on the choice of C, O, Mg, Al, and Na, the adopted nitrogen abundance can affect the value of $\Delta$Y by $\sim$0.01 dex. I conclude that the group of blue-MS stars analyzed in this paper is enhanced in helium by $\Delta$Y=0.08$\pm$0.01 dex with respect to the red MS. 

\subsection{Comparison with the isochrones}
\begin{centering}
\begin{figure*}
 \includegraphics[width=8.7cm]{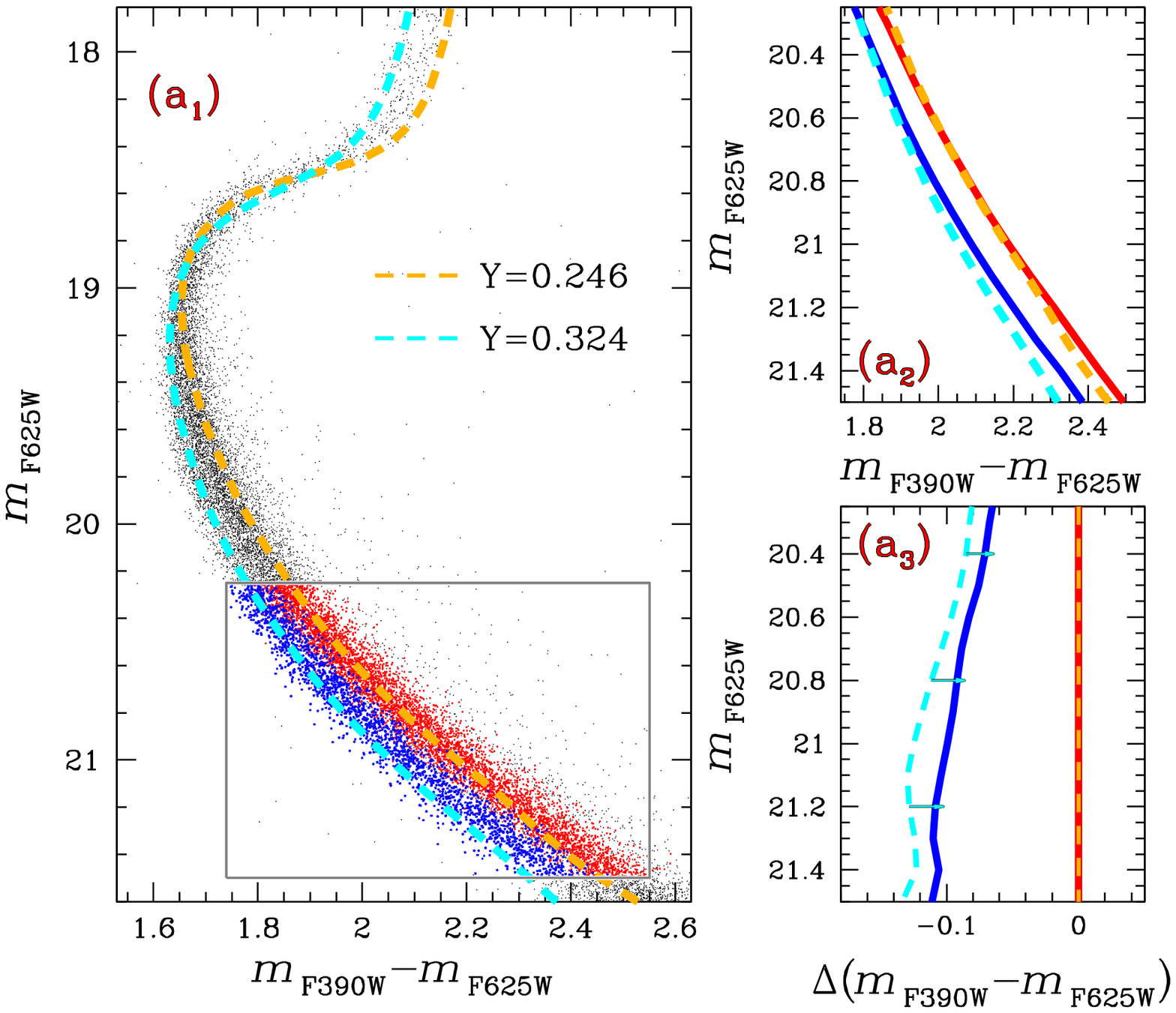}
 \includegraphics[width=8.7cm]{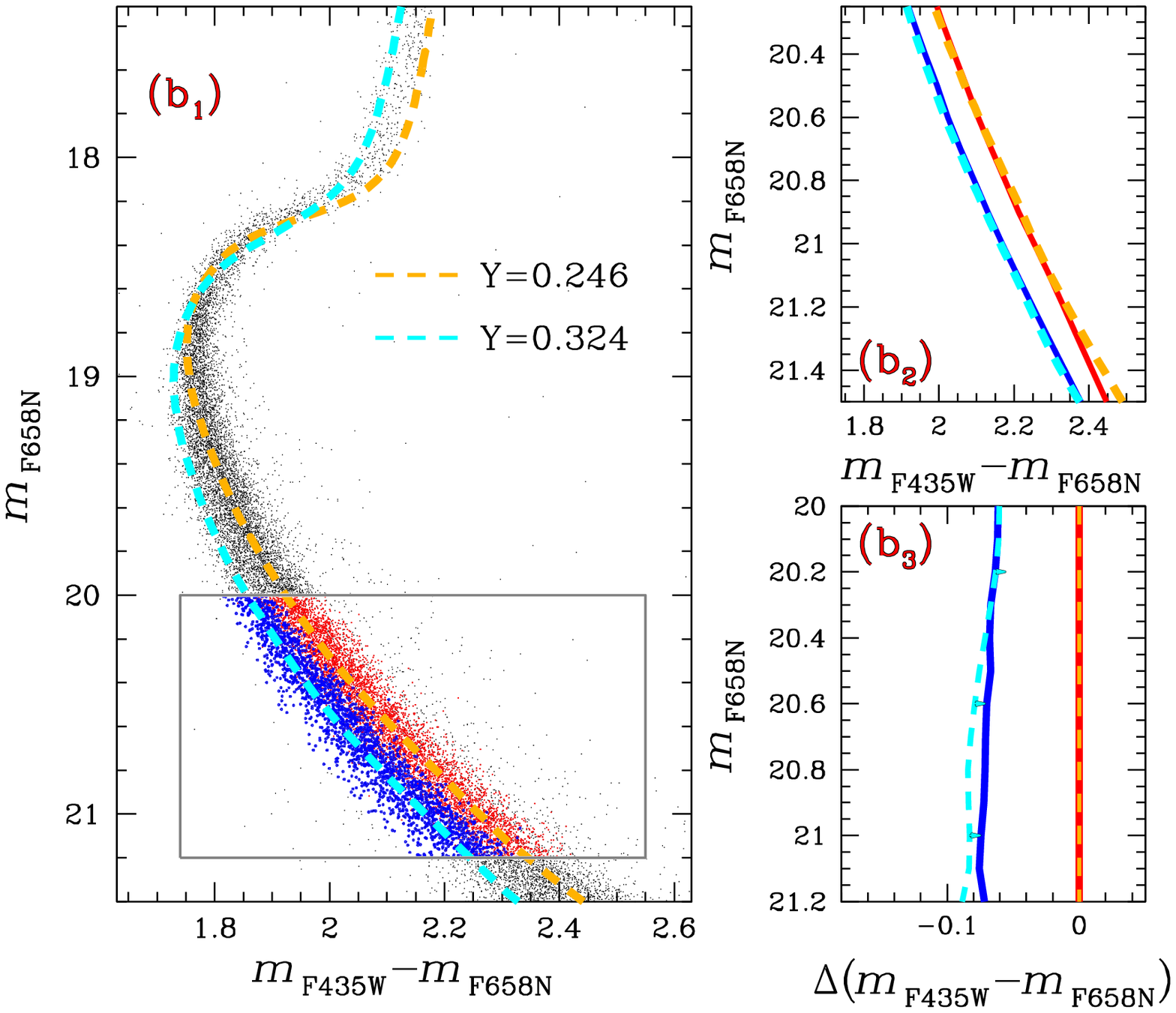}
 \caption{ $m_{\rm F625W}$ vs.\,$m_{\rm F390W}-m_{\rm F625W}$ (panel $a_{1}$) and $m_{\rm F658N}$ vs.\,$m_{\rm F435W}-m_{\rm F658N}$ CMD of stars in NGC\,6626 (panel $b_{1}$). The red- and blue-MS stars selected in Fig.~\ref{fig:sep} are colored red and blue, respectively. The orange and the cyan dashed lines are the best-fitting isochrones with Y=0.246 and Y=0.324 from Dotter et al.\,(2008). The CMDs in the panels ($a_{2}$) and ($b_{2}$) compare the isochrones with the fiducials of the red MS (red-continuous line) and the blue MS (blue-continuous line). The corresponding verticalized CMDs are plotted in panels ($a_{3}$) and ($b_{3}$). The adopted isochrones do not account for the detailed light-elements abundance of multiple stellar populations in NGC\,6266. When the appropriate C, N, O abundance of red MS and blue MS stars are used, the color difference between the two isochrones decreases by the amount indicated by the cyan arrows. See text for details.}
 \label{fig:isocrone}
\end{figure*}
\end{centering}

 The panel ($a_{1}$) of Fig.~\ref{fig:isocrone} reproduces the $m_{\rm F625W}$ vs.\,$m_{\rm F390W}-m_{\rm F625W}$ CMD of Fig.~\ref{fig:fig2}, and the two groups of red-MS and blue-MS stars defined in Fig.~\ref{fig:sep} have been represented with red and blue points, respectively. Two isochrones from Dotter et al.\,(2008) have been superimposed onto the CMD. Both of them have the same iron abundance ([Fe/H]=$-$1.15) and [$\alpha$/Fe]=0.40, but  different helium content of Y=0.246 (orange-dashed line) and Y=0.324 (cyan-dashed line). The helium difference corresponds to that derived from the analysis above. 
I adopted for all the isochrones reddening, E(B$-$V)=0.52, age of 12.5 Gyr, and a distance modulus, $(m-M)_{\rm V}$=15.55 which provide the best fit with the observations.

 The $m_{\rm F625W}$ vs.\,$m_{\rm F390W}-m_{\rm F625W}$ CMD plotted in the panel ($a_{2}$) of Fig.~\ref{fig:isocrone} compares the isochrones with the fiducial lines of the red and the blue MS, which are represented with red and blue continuous lines, respectively. The helium-rich isochrone has bluer colors than the fiducial line of the blue MS in almost all the analyzed interval of magnitude. The isochrone with Y=0.246 is almost superimposed onto the fiducial line of the red MS for $m_{\rm F625W} \lesssim$21.1, but the fit gets worse at fainter magnitudes where the isochrone is slightly bluer than the fiducial.

 In order to do a more thorough and quantitative analysis of the relative helium abundance of stellar populations in NGC\,6266, I have compared in the panel ($a_{3}$) of Fig.~\ref{fig:isocrone} the isochrones and the fiducial lines in the verticalized $m_{\rm F625W}$ vs.\,$\Delta$($m_{\rm F390W}-m_{\rm F625W}$) diagram. To derive the two verticalized fiducials I have subtracted from the color of the red and the blue fiducial, the color of the red fiducial at the corresponding F625W magnitude. Similarly, to determine the verticalized isochrones, I have subtracted from the color of each isochrone, the color of the isochrone with Y=0.246 at the corresponding $m_{\rm F625W}$ luminosity. In this diagram, both the fiducial line of the red MS and the isochrone with Y=0.246 have $\Delta$($m_{\rm F390W}-m_{\rm F625W}$)=0, by construction. 

Panel ($a_{3}$) of Fig.~\ref{fig:isocrone} shows that the verticalized fiducial of the blue MS is redder than the helium-rich isochrone by $\sim$0.02 mag in $m_{\rm F390W}-m_{\rm F625W}$. A small color difference between isochrones and observations is expected, because the adopted isochrones do not account for the  detailed light-element abundance multiple stellar populations in NGC\,6266. 

According to the analysis of Sect.~\ref{sec:elio} the abundance of C, N, O have a significant effect on the $m_{\rm F390W}-m_{\rm F625W}$ color of red- and blue-MS stars.
To estimate it, 
 I have used the procedure described in Sect.~\ref{sec:elio} to calculate two sets of synthetic spectra for blue-MS stars with Y=0.324 at the reference magnitudes $m_{\rm F625W}^{\rm CUT}$= 20.4,  20.8, and 21.2. 
I assumed different chemical-composition mixtures for each set of spectra.
 For the first one, I adopted the same C, N, O abundances used for red-MS stars ([C/Fe]=$-$0.4, [N/Fe]=0.5, [O/Fe]=0.8), and in the second group I used the same chemical composition of the blue MS ([C/Fe]=$-$0.8, [N/Fe]=1.6, [O/Fe]=0.2).    
 Spectra have been used to derive synthetic $m_{\rm F390W}-m_{\rm F625W}$ colors as in Sect.~\ref{sec:elio}. 

 N-rich stars have slightly redder color than N-poor ones, thus suggesting that, when the detailed light-element abundance of the two populations is taken into account, the color difference between the isochrones decreases by $\sim$0.02 mag as indicated by the cyan arrows plotted in the panel ($a_{3}$) of Fig.~\ref{fig:isocrone}.

In the panels ($b_{1}$), ($b_{2}$), and ($b_{3}$) of Fig.~\ref{fig:isocrone}, I have extended the analysis to the $m_{\rm F658N}$ vs.\,$m_{\rm F435W}-m_{\rm F658N}$ CMD. In this case the color distance between the fiducial of the blue MS and the helium-rich isochrone is smaller. Indeed the abundance of C, N, O affects the $m_{\rm F435W}-m_{\rm F658N}$ by $\sim$0.005 mag as indicated by the cyan arrows in the panel ($b_{3}$).
The comparison with isochrones of this section reinforces the conclusion that the two MSs of NGC\,6266 are consistent with two stellar populations with a difference in helium of $\sim$0.08 dex.

Moreover, as discussed in Sect.~\ref{sec:CMD}, the red MS is broadened much more than can be accounted for by photometric errors in $m_{\rm F390W}-m_{\rm F625W}$ by $\sim$0.03 mag. I have used isochrones from Dotter et al.\,(2008) with [Fe/H]=$-$1.15 and different helium to estimate the helium difference that would led to the observed spread and found that, if helium is the only cause of the MS broadening, then it would imply a spread in helium by $\sim$0.025 dex.

\section{Summary and conclusion}
I used photometry from ACS/WFC and WFC3/UVIS on board of {\it HST} to investigate stellar populations in NGC\,6266. 
The CMDs have been corrected for differential reddening, and most of the field stars have been separated from probable cluster members on the basis of stellar proper motions.

The presented photometric analysis
shows, for the first time, that NGC\,6266 exhibits a complex and multiple MS. 
Specifically, a red and blue MS have been detected and found to host the 79$\pm$1\% and the 21$\pm$1\% of MS stars, respectively. In the studied luminosity interval, between $\sim$1 and $\sim$2 F658N mag below the MS turn off, the color separation between the red and the blue MS is about 0.1~mag in $m_{\rm F390W}-m_{\rm F658N}$, and decreases for shorter color baseline like $m_{\rm F435W}-m_{\rm F658N}$ and $m_{\rm F625W}-m_{\rm F658N}$. In contrast, the SGB does not show any significant vertical spread, even at the level of a few hundreds of magnitude. I have demonstrated that the double MS in NGC\,6266 is intrinsic and cannot be due neither to photometric errors nor to residual differential reddening. 
This fact demonstrates that the red and the blue MS of this massive GC correspond to distinct stellar populations. 

Recent work has shown that the analysis of multiple MSs from multi-wavelength photometry allows us to infer precise determination of the helium content of multiple stellar populations. 
To determine the helium content of stellar populations in NGC\,6266, 
synthetic spectra 
assuming different chemical-composition mixtures have been calculated for MS stars, and the 
corresponding predicted colors have been integrated through the {\it HST} filters to allow the comparison with the observed ones. 
 The result of this comparison 
is that the red MS is consistent with 
a stellar population with primordial helium, and O-rich/N-poor, while the blue MS is consistent with being made of He-enhanced, O-poor/N-rich stars.
The enhancement in the helium mass fraction of the blue-MS stars has been derived to be $\Delta$Y=0.08$\pm$0.01. Moreover, the color width of the red MS is much larger that that of blue-MS stars. This additional broadening  could be due by an intrinsic spread in Y  of $\sim$0.025 dex among red-MS stars and suggests that the red MS hosts two or more sub-populations of stars.

 D'Antona et al.\,(2002) first investigated the possibility that stellar populations with different helium abundance may be responsible for the HB morphology of GCs. This group of authors have shown that the HB of NGC\,2808 is consistent with three stellar populations, each with distinct content of helium. They have suggested that the red HB of this cluster is made of stars with primordial (Y$\sim$0.25), while blue-HB stars are helium enhanced up to Y$\sim$0.32 and Y$\sim$0.38, with the hottest HB stars also having the highest helium content (D'Antona \& Caloi 2004, 2008; D'Antona et al.\,2005). This scenario has been confirmed by the discovery by Piotto et al.\,(2007) that NGC\,2808 hosts a triple MS and by the direct measurement of helium abundance from high-resolution spectroscopy of HB stars (Marino et al.\,2014).  

Figure~\ref{fig:red} of this paper as well as Fig.~4 from Piotto et al.\,(2002) shows that NGC\,6266 exhibits an extended HB which is populated on both sides of the RR\,Lyrae instability strip. As suggested by the referee, stars in the bluest tail of the HB of NGC\,6266 could be the precursors of blue-MS stars, in close analogy with what observed in NGC\,2808. Spectroscopy of HB stars together with suitable theoretical HB models (e.g.\,D'Antona et al.\,2005; Busso et al.\,2007; Di Criscienzo et al.\,2014) are mandatory to  firmly establish the location of multiple stellar populations along the HB.

The range of helium abundance variations has been estimated in a few GCs to date. 
So far, helium variations larger than $\Delta$Y$\sim$0.07 have been detected in four massive GCs with extended HB, namely $\omega$\,Centauri, NGC\,6441, NGC\,2808, and NGC\,2419. Less-massive GCs, with less-extended HB, like NGC\,288, NGC\,6397, and NGC\,6752, also have smaller helium variation (e.g.\,King et al.\,2012; Bellini et al.\,2013; Piotto et al.\,2007, 2013; Di Criscienzo et al.\,2010, 2011; Milone et al.\,2012b, 2013).

\begin{centering}
\begin{figure}
 \includegraphics[width=8.5cm]{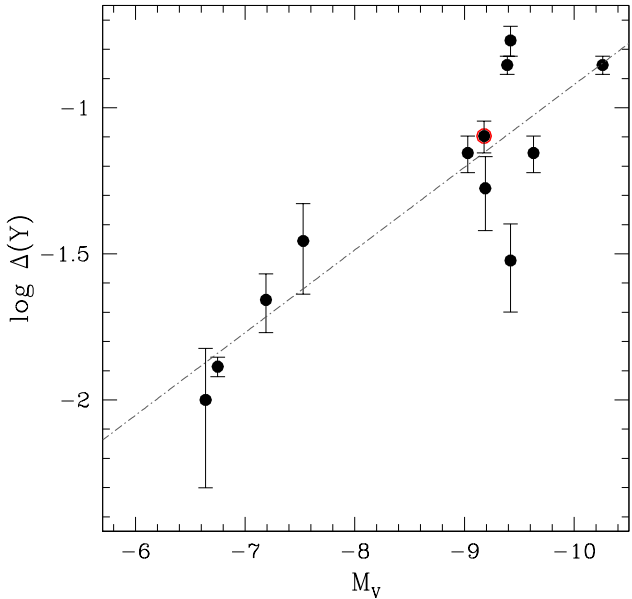}
 \caption{Logarithm of the maximum helium difference among stellar populations in GCs as a function of the GC absolute magnitude. NGC\,6266 is marked with a red circle. The dashed-dotted line is the best-fitting least-squares straight line.}
 \label{fig:elio}
\end{figure}
\end{centering}
 
Figure~\ref{fig:elio} shows the logarithm of the maximum helium difference among stellar populations in GCs based on the analysis of multiple MSs or multiple RGBs, as a function of the absolute magnitude of the host cluster. Black dots mark the eleven GCs studied in literature\footnote{This sample includes the nine GCs listed by Milone et al.\,(2014, see their Table~2), and the values inferred for NGC\,6121 ($\Delta$Y$\sim$0.02, Nardiello et al.\,(2014)) and NGC\,7089 $\Delta$Y$\sim$0.07, Milone et al.\,(in preparation).} while the value for NGC\,6266, inferred here, is marked with a red circle. An inspection of this plot immediately suggests that there is a significant correlation between log$\Delta$Y and $M_{\rm V}$ with a Spearman's rank correlation coefficient $r$=$-$0.70.

 The recent discovery that multiple stellar populations with different helium abundance are a common feature of GCs, provides a new perspective to understand the morphology of their HB and shed light on the second-parameter phenomenon.
 Freeman \& Norris\,(1981) suggested that both a $global$ parameter that varies from cluster to cluster, and a $non-global$ parameter that varies within the cluster, are needed to explain the HB shape in GCs.
 Milone et al.\,(2014) analyzed the HB morphology of 74 GCs observed with ACS/WFC and found that the color distance between the RGB and the reddest part of the HB ($L1$) correlates with the metallicity and age of the hosting GC.
They concluded that metallicity and age are the main $global$ parameters defining the HB morphology in GCs.
 The same authors also found that the color extension of the HB ($L2$) correlates with the cluster mass. The relation between the GC mass and $\Delta$Y  shown in Fig.~\ref{fig:elio} suggests that the range of helium, associated with the presence of multiple stellar populations, is the main $non-global$ parameter. 

The reported results for NGC\,6266, which is among the ten most-massive GCs in the Milky Way and exhibits a very-extended HB, fit very well in the observational scenario for the HB morphology presented in Milone et al.\, (2014). Indeed, the large helium difference between the two MSs of this massive GC further supports the idea of a close relationship among $\Delta$Y, GC mass, and HB extension.

 \section*{acknowledgments}
\small
I acknowledge the financial support from the Australian Research Council through Discovery Project grant DP120100475.
The referee, Paolo Ventura, has greatly improved the quality of this paper. I am also grateful to Anna Fabiola Marino for her work on synthetic spectra and for carefully reading this manuscript. I warmly thank Jay Anderson and his collaborators, who have provided most of the programs for the reduction of {\it HST} data, and Giampaolo Piotto, Luigi Bedin, and Andrea Bellini for many stimulating discussions on GCs and data-reduction techniques.
 I would like to thank Aaron Dotter for the unpublished values of the extinction rate of UVIS/WFC3 filters and Remo Collet and David Yong for useful suggestions on this work.

\bibliographystyle{aa}

\end{document}